# An advanced active quenching circuit for ultra-fast quantum cryptography

**MARIO STIPČEVIĆ,[1,*] BRADLEY G. CHRISTENSEN,[2] PAUL G. KWIAT,[2] AND DANIEL J. GAUTHIER[3]**

[1]*Photonics and Quantum Optics, Center of Excellence for Advanced Materials and Sensing Devices, Ruđer Bošković Institute, Bijenička cesta 54, HR-10000 Zagreb, Croatia*
[2]*University of Illinois at Urbana-Champaign, Department of Physics, 1110 West Green Street, Urbana, IL 61801-3080, USA*
[3]*The Ohio State University, Department of Physics, 191 West Woodruff Ave., Columbus, OH 43210, USA*
**Corresponding author: mario.stipcevic@irb.hr*

**Abstract:** Commercial photon-counting modules based on actively quenched solid-state avalanche photodiode sensors are used in a wide variety of applications. Manufacturers characterize their detectors by specifying a small set of parameters, such as detection efficiency, dead time, dark counts rate, afterpulsing probability and single-photon arrival-time resolution (jitter). However, they usually do not specify the range of conditions over which these parameters are constant or present a sufficient description of the characterization process. In this work, we perform a few novel tests on two commercial detectors and identify an additional set of imperfections that must be specified to sufficiently characterize their behavior. These include rate-dependence of the dead time and jitter, detection delay shift, and "twilighting". We find that these additional non-ideal behaviors can lead to unexpected effects or strong deterioration of the performance of a system using these devices. We explain their origin by an in-depth analysis of the active quenching process. To mitigate the effects of these imperfections, a custom-built detection system is designed using a novel active quenching circuit. Its performance is compared against two commercial detectors in a fast quantum key distribution system with hyper-entangled photons and a random number generator.

## References and links

## 1. Introduction

Detectors of light capable of detecting a single photon, so-called single-photon detectors or single-photon counters, are used in a large variety of scientific research areas and commercial applications, such as quantum information and quantum communication research, light radar (LIDAR), particle counting and sizing, gas analysis, time-resolved spectroscopy, nuclear and particle physics, astronomy, etc. One class of single-photon detector based on a semiconductor material is known as a single-photon avalanche detector (SPAD) operating in Geiger mode capable of counting photons (SPADs). In this paper, we focus on the characterization and performance of silicon-based SPADs, which operate in the visible and near-infrared part of the electromagnetic spectrum.

An ideal single-photon detector generates one logical electrical pulse, typically in the TTL or NIM format, for each photon that hits the SPAD sensor. Realistic detectors, due to the limitations imposed by the physics of the SPAD and imperfections of the electronic circuitry required to amplify and shape the detection signal, deviate from this ideal picture. Manufacturers characterize their single-photon detectors by specifying a set of "standard" parameters, usually limited to detection efficiency, dead time, and dark counts; less often they additionally specify the single photon arrival time resolution (jitter) and the total afterpulsing probability. However, they almost never specify the conditions under which these parameters are valid or sufficiently precisely describe the detection process.

Here, we present an in-depth analysis of the active quenching process that allows us to identify engineering challenges of single-photon detectors based on SPADs. We investigate the range of validity of the "standard" set of parameters for two commercial single-photon detectors. Based on this analysis, we identify an additional set of parameters required to sufficiently characterize behavior of photon counters in demanding applications. These additional imperfections include: rate dependence of the dead time, rate dependence of the photon arrival time resolution (jitter), detection delay shift, twilighting, the temporal distribution of afterpulsing, afterpulsing lifetime, and artifacts of the electronics. We find that these behaviors can lead to unexpected effects or strong deterioration of the detector's performance with respect to what would be expected from the "standard" set only. We then present a novel active avalanche quenching circuit with improved characteristics and discuss its performance in two applications in which single-photon detectors play a major role: beamsplitter-based quantum random number generation and security of an ultra-fast quantum key distribution (QKD) system based on hyper-entangled photons.

## 2. Active quenching process

SPADs can be divided into two broad categories: thick structures (also known as “reach-through”) and structures with shallow absorbing layers [1]. The main tradeoff in these two types is between the detection efficiency and jitter, but they both require quenching.

In a SPAD reverse-biased above its Geiger breakdown voltage, a single absorbed photon can trigger a macroscopic, self-sustaining avalanche current. Once started, an avalanche may last indefinitely and thus quenching must be used to ensure that another photon can be detected. Quenching can be done by quickly lowering the bias voltage below the Geiger threshold and restoring it after a predetermined hold-off time. This can be achieved passively by passing the avalanche current a series resistor, but is best achieved by means of active electronics switches. During an avalanche, the charge multiplication process generates a sizeable current that is comfortably above the readout noise of the subsequent amplifying stage and thus a single photon can be detected reliably. A typical time variation of the bias voltage across the SPAD, during one active quenching instance is shown in Fig. 1.

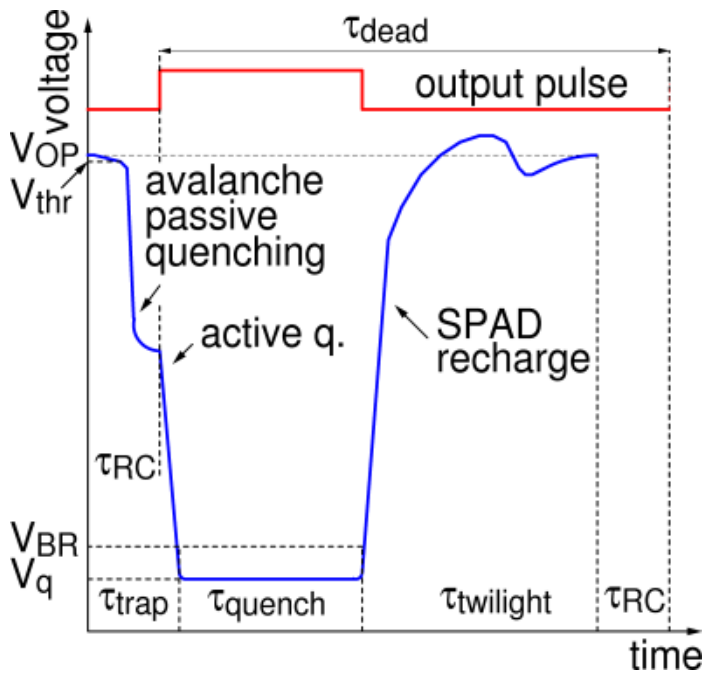

**Fig. 1.** Typical temporal evolution of the voltage across a SPAD during an active avalanche quenching event. Here, the bias voltage across SPAD during a quenching sequence is shown in blue, $\tau_{\mathrm{trap}}$ is the interval during which deep states (traps) are filled, $\tau_{\mathrm{quench}}$ is the interval during which the SPAD is biased below the Geiger threshold, $\tau_{\mathrm{twilight}}$ is the interval during which SPAD is (partially) sensitive to single photons but the electronic circuit amplifying stage is shut down, $\tau_{\mathrm{RC}}$ is the propagation delay of the quenching stage, $\tau_{\mathrm{dead}}$ is the dead time, $V_{\mathrm{OP}}$ is the nominal operating voltage of the SPAD, $V_{\mathrm{thr}}$ is the pulse discriminating level, and $V_{\mathrm{BR}}$ is the Geiger breakdown voltage of the SPAD.

Because of the time course of the detection process, there is a temporal window in which photons cannot be detected; that is, the detector must have a certain "dead time" $\tau_{\mathrm{dead}}$, commonly defined as the minimum time between two consecutive photon detections. It is usually defined by internal electronics delays and therefore quite stable under various test conditions. The dead time of a single-photon detector can be measured as the shortest observed waiting time between photon detections, as shown in Fig. 2 or Fig. 10(a).

It is widely taken that dead time is equal to the photon pair resolution time. There is perhaps a historical reason for this belief because the pulse pair resolution is identical to the dead time for photomultiplier detectors where quenching is not necessary, but the same does not hold for SPAD-based photon counters. This is due to the fact that restoring the bias voltage across the SPAD after quenching can only be done in a finite time, and may feature a few ripples before settling at the nominal value as shown in Fig. 1; therefore, in general, the pulse pair resolution may be slightly longer than the dead time. To avoid retriggering of the quenching circuit, it is necessary to keep the sensing electronics shut off until the bias voltage is settled and the SPAD begins to operate at its nominal bias voltage. The part of the dead time between the end of quench and start of sensitivity of the sensing amplifier, denoted by $\tau_{\mathrm{twilight}}$ in Fig. 1, is called the "twilight zone" [2]. During this interval, the SPAD is at least partially sensitive to incoming photons and can go into avalanche even though the detector is technically in the dead time. If this happens, the detector will generate an output pulse immediately after the end of the dead time; a process named "twilighting."

Finally, the detection jitter has three contributions: variation due to the spread of drift times from the place of photon conversion to the avalanche region; statistical fluctuation of the avalanche current [3]; and imperfect settling of all voltages in the circuit between subsequent detections, which generally tends to worsen the jitter as the detection rate increases. We will discuss jitter in greater detail in Sec. 4.

## 3. Imperfections in commercial single-photon detectors based on SPADs

The "standard" imperfections mentioned above namely, non-unity detection efficiency, dead time, dark counts, afterpulsing and jitter, define the most common parameters used for characterization of single-photon detectors. However, we find that these parameters do not sufficiently characterize the detection processes and that some other imperfections exist that sometimes must be taken into account.

We start by studying the "standard" imperfections of the well-known photon-counting module SPCM-AQRH (PerkinElmer, manufactured in 2008 before the electronics was redesigned by Excelitas who took over this PerkinElmer product line). In order to characterize the device performance, we measure a large number of time intervals between consecutive output pulses when the detector is subjected to weak incoherent light from a light emitting diode (Hamamatsu L7868, wavelength 670 nm, spectral width 30 nm full-width-at-half-maximum (FWHM)) operated at a constant optical power, generating about 50,000 counts per second (dark counts and afterpulses included). The time intervals are measured by a high-resolution (17.7 ps FWHM) time-to-amplitude (TAC) converter (Ortec TAC model 567) whose analog output is digitized with a 16-bit digital-to-analog converter (National Instruments model USB-6251) connected to a LabVIEW program running on a personal computer. The intervals are displayed in a histogram shown in Fig. 2.

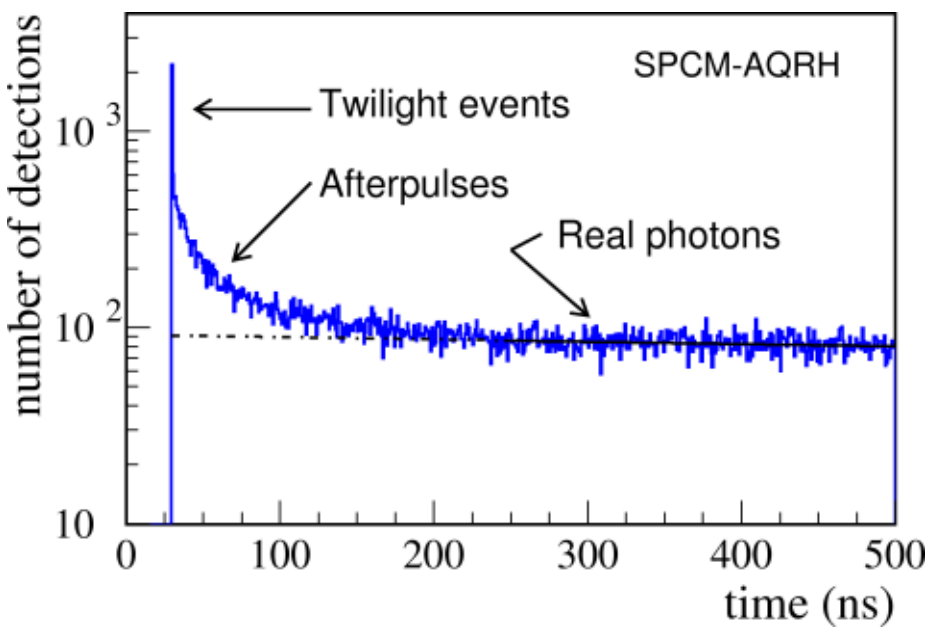

**Fig. 2.** Histogram of time intervals between subsequent electrical pulses recorded from the detector SPCM-AQRH illuminated by a constant-power LED to achieve total count rate of 50

kcps. Histogram bins are 1-ns wide. The data collection time for the 61638 events shown in the histogram is ~3 minutes.

Two types of events are clearly discernible: (1) detection of real photons, which follows an exponential probability distribution [4]; and (2) afterpulses that originate from traps in the SPAD [5]. The afterpulsing probability and the leading trap lifetime, obtained using the deep level spectroscopy described in [5], are $(0.68 \pm 0.04)\%$ and $(32 \pm 2)$ ns, respectively. The dead time is visible in Fig. 2 as a gap between zero and the onset of high detection probability, which is $\tau_{\text{dead}} = (29.1 \pm 0.1)$ ns for the data shown here. We use the same method to determine dead times of other detectors in this study. The dark counts rate, determined in a separate measurement with no light, is $(726 \pm 10)$ cps. All these parameters are in agreement with the datasheet, with exception of the trap lifetime, which is not specified. Finally, according to the datasheet, the detection efficiency at 670 nm is about 65%. This concludes a list of standard parameters. However, in Fig. 2 we also observe a strong excess of events appearing in the first 1-ns temporal bin just after the dead time. These events probably correspond to the twilighting effect explained above. In order to study in detail this and other non-standard imperfections, we use the setup shown in Fig. 3.

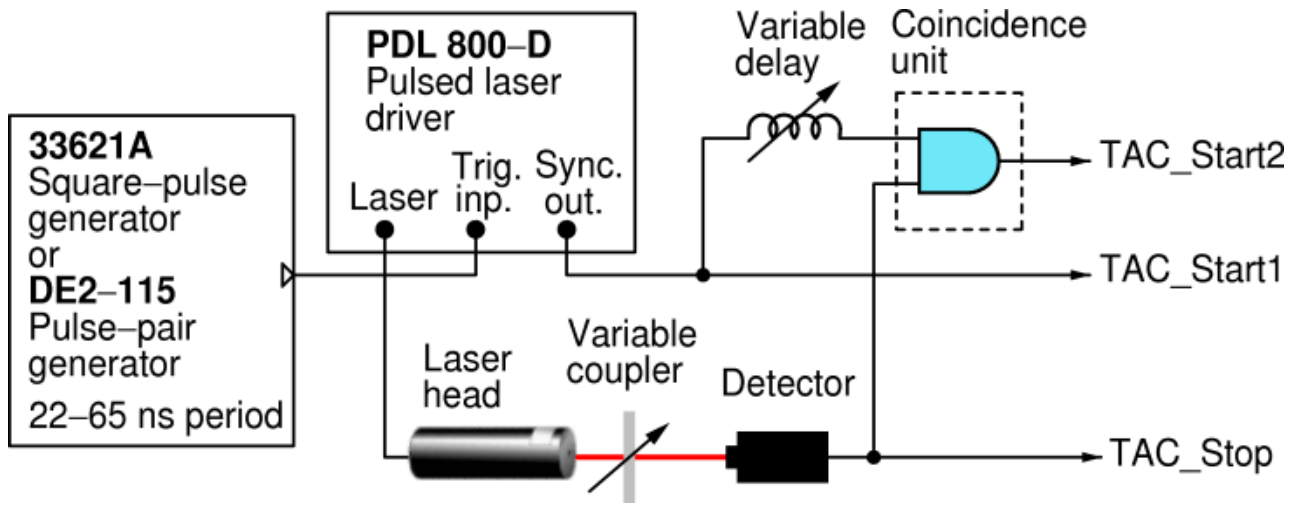

**Fig. 3.** Setup for characterizing jitter, detection shift, and twilighting of photon detectors.

The picosecond pulsed laser (PicoQuant) consists of the driver (PDL 800-D) and the laser head (LDH-P-670) featuring an optical pulse width of 39 ps FWHM at wavelength of 676 nm. An optical pulse is generated by applying a TTL pulse at the trigger input (Trig. Inp.). The synchronization output (Sync. Out.) provides a logic pulse in synchronization with the optical pulse. The laser is coupled to the detector being tested via a manually adjustable coupler control of the detection rate. The setup has 3 digital outputs (TAC_Start1, TAC_Start2, TAC_Stop) that are used to start and stop the TAC time measuring system mentioned above, as will be explained in greater detail below. The coincidence unit generates a 10-ns-long TTL pulse if a pulse from the detector under test and a laser pulse happen together within a 5.0-ns temporal window. We perform two types of experiments with this setup.

The first measurement consists in determining the time interval between emission of a light pulse and its detection with a periodic generation period of 30 ns. We measure two parameters as functions of the detection rate: the average time interval and its spread (FWHM). Both parameters are measured using the TAC_Start1 and TAC_Stop signals; the resulting distributions for 4 detection rates are shown in Fig. 4. We note that the photon arrival-time resolution worsens from 335 ps at low detection rates (up to ~100 kcps), all the way to 608 ps at a 4 Mcps detection rate. At the same time, the peak of the photon detection time delay distribution shifts by 855 ps towards longer times. The jitter, jitter degradation, and the detection time shift are not specified in the datasheet [6].

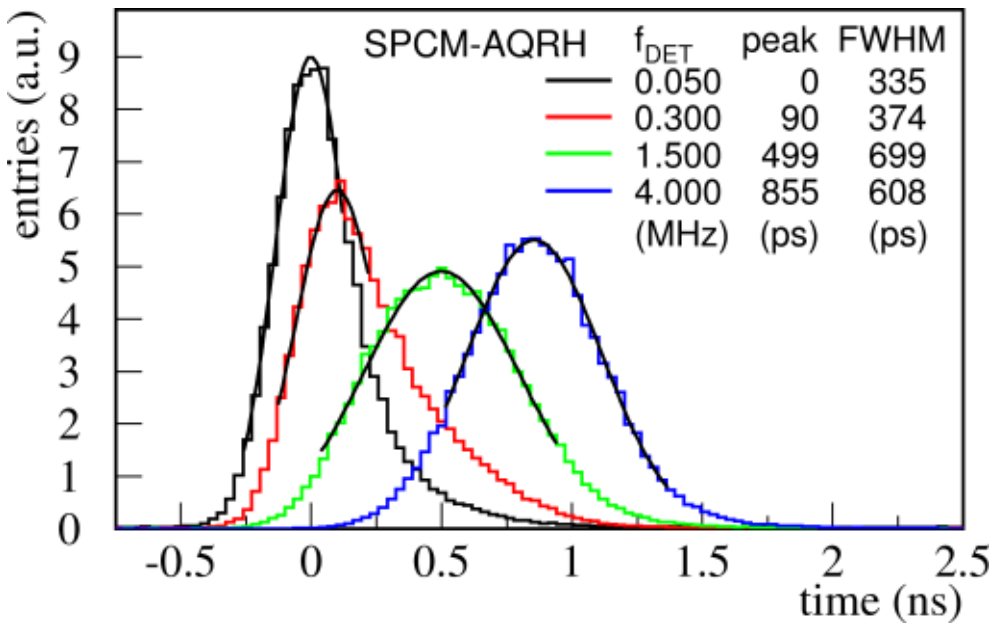

**Fig. 4.** Histogram of jitter of the SPCM-AQRH detector as a function of the mean detection rate for counting rates of a) 50 kcps, b) 300 kcps, c) 1.5 Mcps, and d) 4.0 Mcps. The delay between photon emission and detection is on the abscissa, while the number of events is on the ordinate. The inset table lists the detection rate ($f_{\mathrm{DET}}$), peak photon detection time (peak), and jitter (FWHM) obtained by a Gaussian fit.

The second measurement allows us to investigate the twilighting effect in two commercial detectors: the SPCM-AQRH and a SPD-050 (Micro Photon Devices, MPD) [7]. To that end, we illuminate a detector with a pair of weak consecutive laser pulses separated by $\Delta T$, accomplished using a computer-controlled custom-built electronic pulse pair generator (Terasic FPGA DE2-115), which triggers the picosecond laser. The delay $\Delta T$ between two laser pulses can be set in the range $10-255$ ns in steps of 1 ns. The pair repetition period is fixed at 1 μs, much longer than the afterpulsing lifetime, such that the afterpulses die off completely between subsequent pairs. The detection probability of a laser pulse is set to about 0.05 so that each pulse wavepacket incident on the detector essentially contains either a single photon or no photon given the detector efficiencies at the laser wavelength are relatively high (65% for SPCM-AQRH, 33% for SPD-050). The appearance of a logic pulse at the TAC_Start2 output in the setup in Fig. 3 indicates that a first photon in the pair has been detected. Then, the time until the next detected photon (TAC_Stop) is measured by the TAC, and a histogram is generated. From these data, for each pulse delay $\Delta T$, we calculate three parameters: (1) the detection probability of the second photon in a pair relative to the detection probability of the first photon when the first photon is detected; (2) the average time delay between detections of the first and second photons; and (3) the jitter of the time delay between detections of the first and second photons.

Figure 5 shows twilighting for the two detectors. The dead time of each detector is indicated by a vertical line. We see that, for both detectors, the probability of detecting the second photon goes from zero for small $\Delta T$, to near unity before $\Delta T$ reaches $\tau_{\mathrm{dead}}$. In particular, SPD-050 has two outputs: a TTL-level compatible "TTL output" and a NIM-level compatible "Timing output". Not reported in the datasheet, we find that the Timing output has significantly shorter twilight zone and a bit shorter dead time. Both imperfections play an important role in QKD, as will be discussed in Sec. 5.

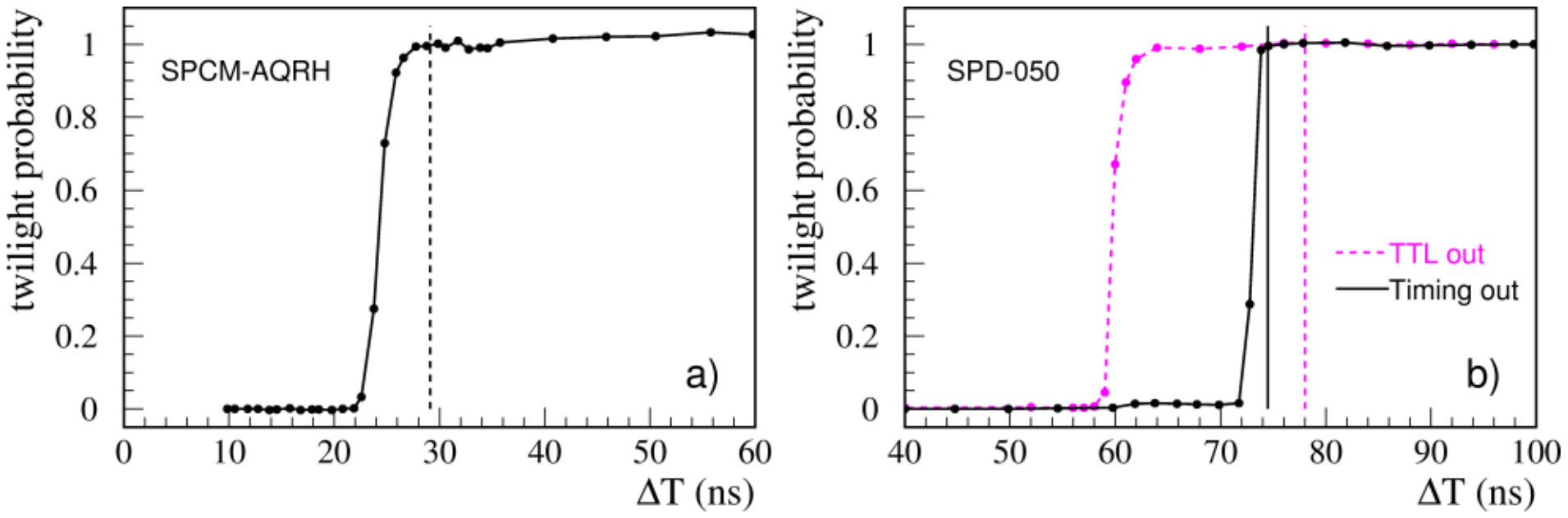

**Fig. 5.** Twilighting effect in the a) SPCM-AQRH and b) SPD-050 (TTL output and Timing output) detectors. The detection efficiency of the second photon in a pair, relative to the efficiency of the first photon (when the first photon is detected) as a function of the delay between the two photons. The dead times of a) 29.1 ns and b) 78.0 ns for TTL output and 74.5 ns for Timing output are indicated by the vertical lines.

We observed that, in both detectors, the electrical output pulse corresponding to the second photon, the one detected in the twilight zone, appears just *after* the detector dead time. This means that a photon detected in the twilight zone appears *shifted* in time by as much as the duration of the twilight zone. This temporal shift may give rise to considerable problems in applications sensitive to photon timing in which photons can appear in the twilight zone of a detector.

To illustrate this effect, the average time delay between detection of the first and the second photon is shown in Fig. 6(a) for the detector SPCM-AQRH, where it is seen that some shift persists for as long as $15 - 20$ ns after the dead time, which we believe is due to a temporary rise of the avalanche sensing threshold voltage $V_{\text{thr}}$ (explained in Fig. 1) following a photon detection. The detection shift falls below our systematic error ($\sim 100$ ps), for a photon pair delay $\Delta T > 50$ ns. We defer the discussion of this behavior for the SPD-050 until later.

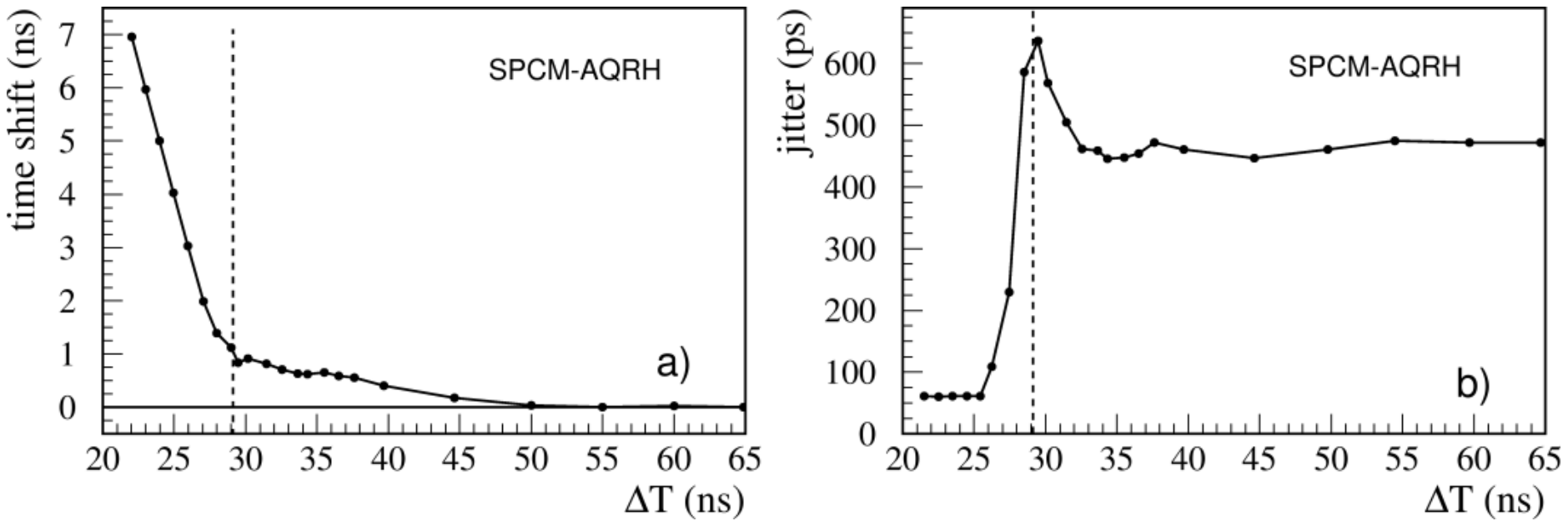

**Fig. 6.** a) Time shift between the true and measured photon arrival time for the second photon in a pair (if both photons have been detected), as a function of the time interval between the two incoming photons. b) Time resolution (jitter) FWHM of the second photon in a pair if both photons have been detected. Dotted lines mark respective dead times.

We further investigate whether the time resolution (jitter) is affected by the temporal separation $\Delta T$ between two photons. To that end, we consider only those events in which both photons in a pair are detected, and measure the jitter between them as a function $\Delta T$. Results for the SPCM-AQRH are shown in Fig. 6(b). To understand this behavior, let us first consider $\Delta T$ large enough that all voltage transitions in the quenching circuit are settled down between subsequent detections, which certainly is the case when the average detection frequency is about 100 kcps (50 kcps per laser pulse), thus the jitter should correspond to the low-rate value observed in Fig. 4 ($335$ ps FWHM). Because we measure the time between two independent photon detection events, the observed jitter should be about $\sqrt{2} \cdot 335 \approx 472$ ps FWHM, which is indeed close to the value found in Fig. 6(b) for large $\Delta T$. As $\Delta T$ becomes smaller, the jitter rises, reaching its peak value at $\Delta T = \tau_{\text{dead}}$ . This may be due to an incomplete recovery of the bias voltage across the SPAD between the two detections, causing it to operate at a lower than nominal voltage. Finally, photons detected in the twilight zone ($\Delta T < \tau_{\text{dead}}$ ) appear to have much smaller jitter, but this is an illusion because we are essentially measuring the length of the dead time in this case. The true uncertainty in the temporal detection of these photons extends over the entire twilight zone and the observed small jitter corresponds only to the delayed electronic detection pulse.

The timing output of the SPD-050 detector has a substantially shorter twilight zone than the SPCM-AQRH detector and it does not suffer the jitter and shift variations when $\Delta T >$ $\tau_{\text{dead}}$ (data not shown). We hypothesize that the overall better performance of this detector beyond $\tau_{\text{dead}}$ is due to the fact that it is made with special care to achieve the best timing resolution in the visible range, (typically between $35 - 50\text{ps}$ FWHM at its timing output), and very low afterpulsing, (typically < 0.5%) [7]. For both of these goals, a prolonged quenching period much longer than $\tau_{\text{dead}}$ and sufficient time for bias settling seem necessary and effectively mitigate the post-dead-time non-idealities observed in Fig. 6 for the SPCM-AQRH detector.

## 4. An improved custom-built actively quenched detector

While effects of the imperfections discussed in the previous section may be negligible for some applications, for other applications they may be severe or prohibitive, as will be discussed in Sec. 5. To reduce the effects of these non-idealities, we designed an improved single-photon detector module based on a novel active quenching circuit optimized for our fast QKD system.

We identified the SAP500 SPAD from Laser Components [6] as the best candidate for our purposes. SAP500 has a reach-through structure illuminated from the back side, sensitive in the visible (VIS) and the near-infrared (NIR) wavelength spectral regions [8]. A broad-band anti-reflective (AR) coating on the entrance surface minimizes reflectance for the incoming photons. A photon entering through the bottom side may convert into a single free charge carrier in the conversion region ($\pi$-region) situated between the p+ layer and the p+ region, as shown in Fig. 7. In the Geiger mode, a single-photon detection starts by converting the photon into a single charge carrier in the low-doped conversion region ($\pi$ - region). In this region, the electric field is low, enabling the charge carrier to quickly (in about 10 ps) reach a saturated velocity and to drift into the PN region (typically in about 100 ps). In the PN region, consisting of p+ and n+ regions, the electric field is sufficiently strong to allow the carrier to acquire enough energy to generate a self-sustaining cascade (avalanche). The conversion region in the SAP500 is only about 25 µm thick, which is on the thinner side for a reach-through structure. However, a special characteristic of this SPAD is that parts of the bottom and top surfaces are covered with a metal layer that acts as a mirror and effectively increases the length of the available photon-conversion path. This improves the QE in the long-wavelength end of the sensitive spectrum, while at the same time enables good photon arrival timing performance (low jitter). Yet another benefit of the short conversion region is the possibility of operating at low $V_{BR}$ (~125 V at 22 °C junction temperature, while other SPADs with similar spectral QE typically have $V_{BR}$~250-500 V).

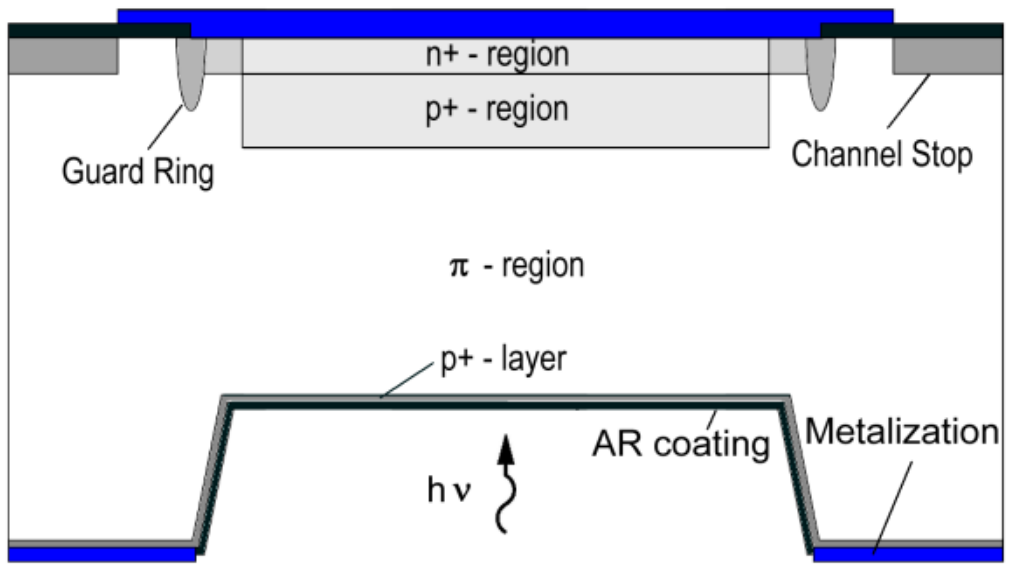

**Fig. 7.** Illustration of the cross section of the SPAD SAP500, manufactured by Laser Components GmbH. The active area is 0.5 mm in diameter (not drawn proportionally). The photon conversion region is situated between the $p^+$ layer and $p^+$ region, and is typically 25 µm thick.The bottom and the contact part of the top side are covered by metalized layers whose purpose is to enhance containment of a photon and thus its conversion to a free carrier.

While the efficiency of the photon conversion, termed "quantum efficiency" (QE), is essentially a property of the material composition and is constant for a given wavelength, the probability of avalanching depends strongly on the bias voltage above the Geiger threshold $V_{\mathrm{BR}}$, the so-called the "excess voltage" $V_{\mathrm{E}}$. The avalanching probability is zero when $V_{\mathrm{E}} \leq 0$ and it rises with $V_{\mathrm{E}} > 0$ to a maximum value which, together with other inefficiencies, determines the overall detection efficiency. Since the dark count rate also rises with $V_{\mathrm{E}}$, there is a practical limit or an optimum value for the excess voltage, which depends on the intended application.

In our detector, we use SAP500-T8 device, which encloses in the industry-standard TO-8 package the SAP500 chip mounted on top of a double-stage thermoelectric cooler (TEC) coupled to a negative-temperature-coefficient sensing resistor (NTC). We use the TEC/NTC combination to stabilize the SPAD's junction temperature to -10 $^{\circ}$C, which reduces the dark counts, but increases the afterpulsing probability; this choice of temperature balances these two characteristics against the desired application. The detection efficiency, obtained with our avalanche quenching circuit described below, measured relative to the known efficiency of SPCM-AQRH at an excess voltage $V_{\mathrm{E}} = 15$ V, is $(73 \pm 2)\%$ at 710 nm.

The design of our custom-built avalanche quenching (AQ) circuit is shown in Fig. 8(a). A positive high voltage (HV), supplied by a miniature DC-to-DC converter (EMCO Q-series), supplies the bias voltage for the SPAD. This AQ circuit follows the general quenching sequence shown in Fig. 1 and is similar to our previous design [9], with the crucial difference that the capacitive coupling between the quenching signal and the SPAD is replaced with galvanic coupling. Namely, while capacitive coupling offers a simpler circuit, it has a problem that the coupling capacitor has less time to completely charge and discharge as the average detection frequency increases, which leads to a lower quenching voltage step ($V_{\mathrm{step}} = V_{\mathrm{OP}} - V_{\mathrm{q}}$) as well as to a shift of the avalanche sensing threshold level ($V_{\mathrm{thr}}$). These effects lead to an elongation of the dead time, higher jitter, and shift in the detection delay towards longer times. The last two imperfections are also seen in the SPCM-AQRH, as shown in Fig. 4. The jitter degradation and timing shift in the SPCM-AQRH module have been observed previously [10], where an improved design of the readout electronics based on capacitive coupling of the quench voltage pulse was proposed.

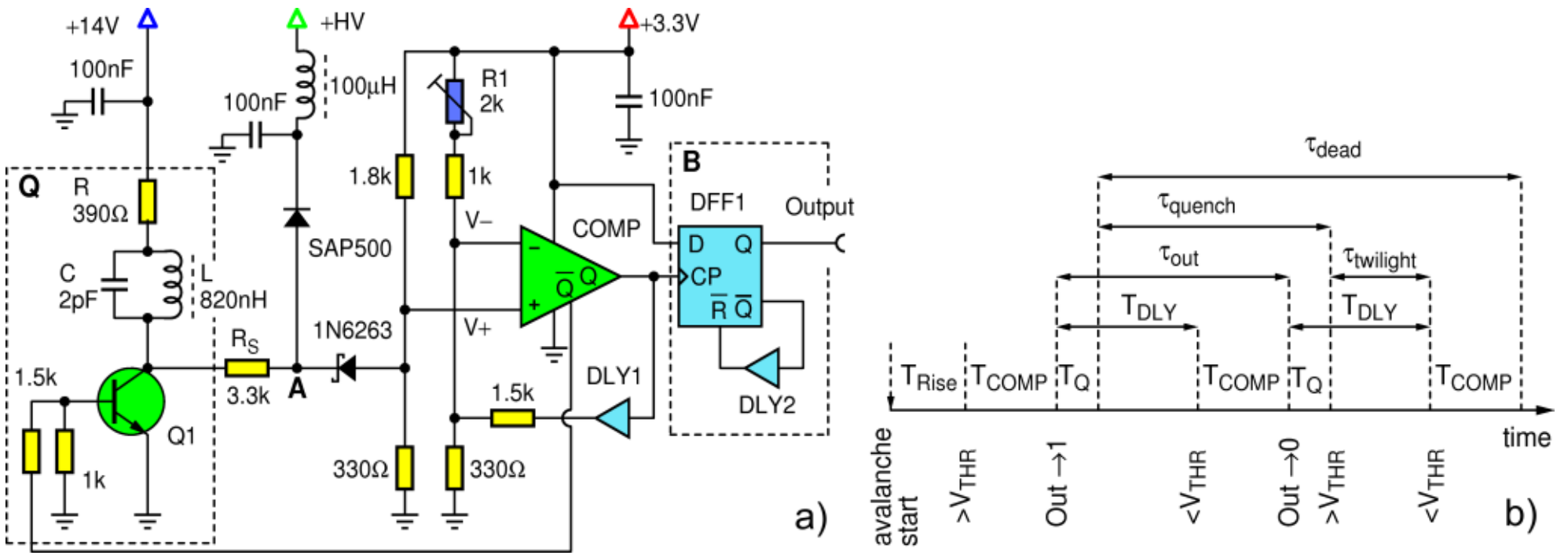

**Fig. 8.** a) Schematic diagram of the improved avalanche quenching circuit. COMP is a fast comparator AD8611 (Analog Devices). b) Timing diagram of the photon-detection cycle of the avalanche quenching circuit.

The detailed timing analysis of the quenching process of this circuit is given in Fig. 8(b). An avalanche in the SPAD causes a rapid rise in voltage at point **A** and consequently at the positive input of the comparator COMP. When the signal-sensing threshold level at the positive input (determined by the value of the resistor $R_1$) is surpassed, the comparator goes into the HIGH logic state, causing the quenching pulse stage (**Q**) to generate a jump of $V_{\mathrm{step}} = 25$ V at the point A, enough to quench the SAP500 operation at an excess voltage of up to 20 V, as shown in Fig. 9. In order to achieve fast transition edges (about 2-ns rise and 4-ns fall time) we use a critically damped resonant circuit (RLC) as a collector load of transistor Q1. Electromagnetic energy stored in the inductance (L) enables the voltage across the RLC circuit to temporarily jump to the value higher than the supply voltage (14V), thus enhancing the quenching capability. The output pulse is delayed by about 9 ns from the quenching pulse due to the propagation times through COMP and the blanking circuit.

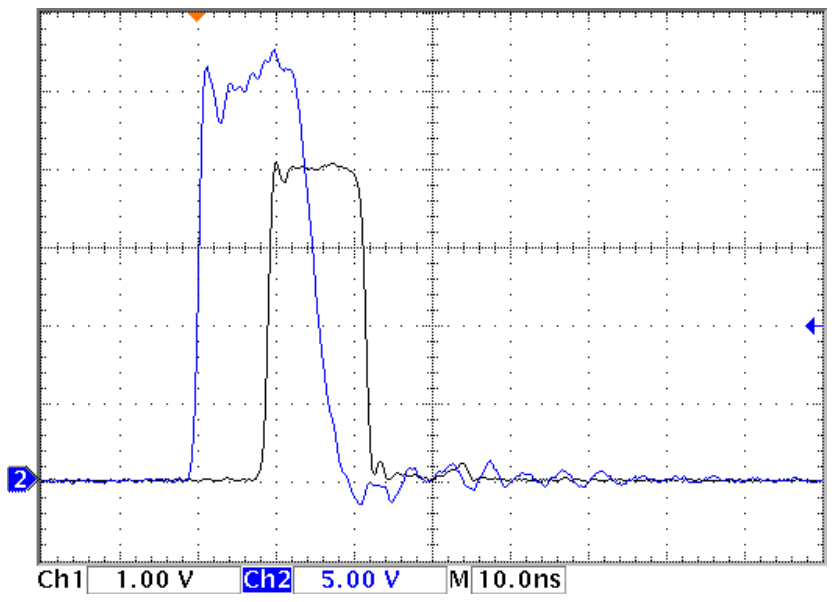

**Fig. 9.** Waveform of a quench pulse at point A (blue curve) and of the output pulse (black) (color online).

During the combined propagation delay through the COMP and the stage **Q** (about 5 ns in total) the SPAD is passively quenched by means of the series resistor $R_S$, after which active quenching takes over (Fig. 1) and the circuit is locked in this state. Later, the HIGH state of COMP, delayed by $T_{DLY\,1}$ through the delay stage DLY1, raises the signal-sensing threshold level beyond the signal level, thus causing COMP to go LOW and the voltage at point A to zero. This ends the active quenching of the SPAD, but does not terminate the latched state of the entire quenching loop. The remaining part of the loop is the twilight zone during which the LOW of the COMP state propagates through DLY1, restoring the sensitivity of the comparator to the next avalanche. Because the SPAD is partially or fully biased during this period, it can go into an avalanche. However, the output pulse will only appear after $\tau_{dead}$ . Values of the timing parameters, measured at a low detection rate ($\approx$ 1 Mcps) using an oscilloscope probe with resistance of 1 M and capacitance of 0.9 pF, are: the time from the avalanche until the comparator is triggered $T_{rise} = 0.5$ ns; the comparator propagation delay $T_{COMP} = 4.5$ ns; stage propagation delay $T_Q = 0.5$ ns; and buffer delay $T_{DLY\,1} = 6$ ns. From these experimentally determined values, one can calculate the other parameters:

$$T_{twilight} = T_{DLY\,1} - T_Q = 5.5 \text{ ns} \quad (1)$$

$$T_{quench} = T_{DLY\,1} + T_{COMP} = 10.5 \text{ ns} \quad (2)$$

$$\tau_{dead} = 2(T_{DLY\,1} + T_{COMP}) - T_Q = 21.5 \text{ ns.} \quad (3)$$

The fast turn-on and turn-off of our optimized quenching pulse allow for a relatively short twilight zone whose duration is determined by the propagation delay time $T_{DLY\,1}$ of the buffer DLY1.

We now discuss of the various design choices in our quenching circuit that affect the detection module performance. For our QKD protocol (described in Sec. 5), the detectors should be optimized for a short dead time, stable detection delay (with minimal twilighting), high detection efficiency and low jitter. Due to the use of a small coincidence detection window, dark counts up to the level of a few kcps and afterpulse probability up to a few percent can be tolerated.

The short dead time is achieved in our design by using a fast comparator and a tight timing sequence of the two feedback loops. However, we observe that when the detector is illuminated by continuous-wave light with Poisson statistics, the dead time $\tau_{dead}$ elongates from 21.5 ns at low detection rate to about 23.5 ns at a rate of 30 MHz, probably due to the effects of unavoidable parasitic capacitances in the AQ circuit. To eliminate this effect and to reduce the twilight interval, the output of the detector is filtered through a blanking circuit (**B**), as shown in Fig. 8(a). The blanking circuit normally transmits pulses from its input to its output but withholds any pulse that appears at a time shorter than $T_{\mathrm{B}} = 2\,T_{\mathrm{DLY\,2}}$ after the previous transmitted pulse. To achieve this function, we use a non-retriggerable monostable multivibrator circuit [11]. We chose $T_{\mathrm{B}} = 24$ ns, just a bit longer than the longest observed dead time, so it becomes the new, stable dead time, and that reshapes the output pulse width to 12 ns, as can be seen from the direct measurement shown in Fig. 10(a). At the same time, the twilight interval is reduced to less than 1.5 ns, as shown in Fig. 10(b).

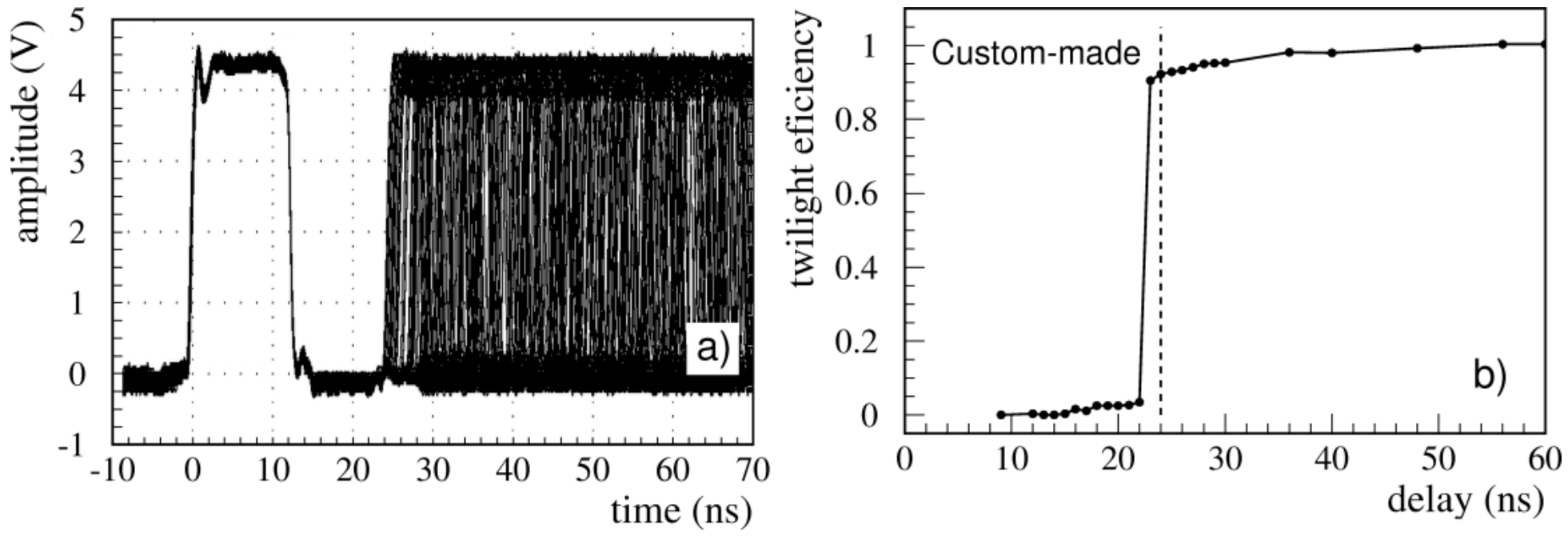

**Fig. 10.** a) The cumulative output of a single-photon detector module illuminated by Poissonian light from a light emitting diode (LED) attenuated to yield approximately 50,000 counts per second on average. The 24-ns dead time is defined as the minimum time delay between the trigger event on the left and the next pulse. b) Twilighting. The dead time of 24 ns is indicated by the vertical dashed line.

The detection time shift and jitter as functions of the detection rate, when the detector is illuminated with laser pulses at a repetition rate with $T = 30$ ns (the first type measurement described in Sec. 3), are shown in Fig. 11(a). The jitter varies from 164 ps (below 300 kcps) to 233 ps (at 4 Mcps) and the systematic detection time shift is only about 26 ps over the whole detection range from $0 - 4$ Mcps. Yet another way to evaluate the detection time shift and jitter is testing with pulse pairs, where the temporal separation between the pulses is $\Delta T$ (second type measurement described in Sec. 3), as shown in Figs. 11(b) and 11(c). The good stability of jitter performance, stable detection delay and virtual absence of twilighting are enabled by a fast bias voltage restore shown in Fig. 9 and short settling times within the AQ circuit.

We see that our custom-built detector outperforms the SPCM-AQRH detector module in a couple of standard imperfections. It has smaller dead time (24 ns as opposed to 29.2 ns for our SPCM-AQRH sample and up to 50 ns according to the datasheet) and smaller jitter (164 ps as opposed to 350 ps at low detection frequency and 233 ps as opposed to 608 ps at 4 Mcps). It is also superior in several non-standard imperfections: less twilighting (compare Fig. 10(b) to Fig. 5(a)), smaller detection delay shift (26 ps compared to 855 ps) and better jitter stability (compare Figs. 11(a), 11(b) and 11(c) to Figs. 4, 6(a) and 6(b), respectively).

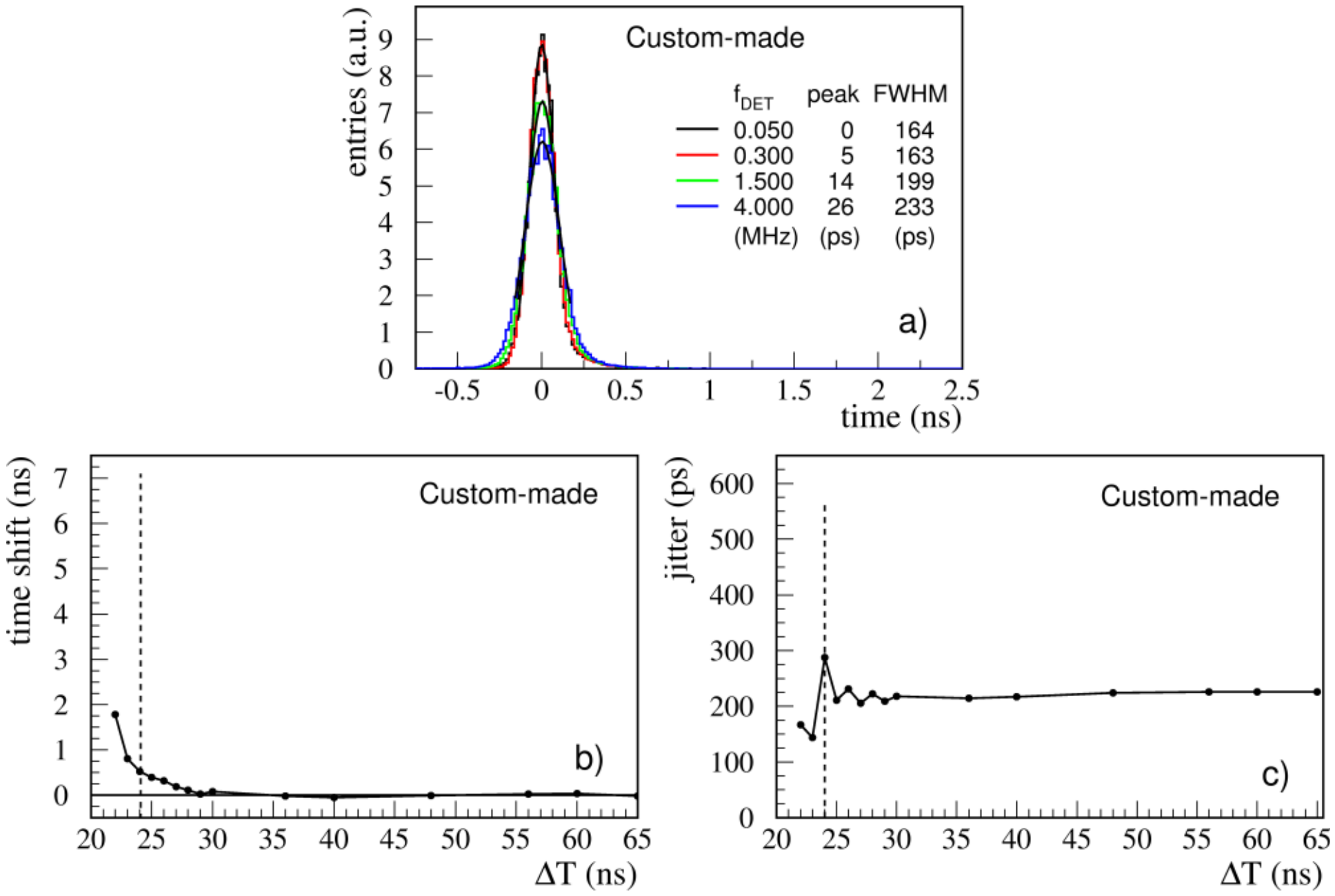

**Fig. 11.** a) Histogram of jitter timing of our custom-made detector as a function of the mean detection rate, when illuminated with laser pulses at repetition rate with $\Delta T = 30$ ns. The delay between photon emission and its detection is on the abscissa, while the number of events is on the ordinate. The inset table relates detection rate ($f_{\mathrm{DET}}$), peak photon detection time (peak), and jitter (FWHM) obtained by a Gaussian fit. b) detection time shift, and c) jitter, for the custom-built detector. The dead time of 24 ns is indicated by the vertical dashed lines. Axes spans of the plots b) and c) are the same as in Fig. 6(b) and Fig. 6(c) respectively, for easier comparison.

Afterpulsing in a silicon (Si) SPAD is usually well explained by deep trap states having a unique lifetime, which leads to a decaying exponential probability distribution function (p.d.f) of afterpulses, and this has been found to be a good approximation for SAP500 [4]. Also, Si SPADs with two or more distinct lifetimes have been observed [5], [12]. Similarly, afterpulses are even more prevalent in InGaAs/InP SPADs, in which lifetimes of the deep trap states form a continuum such that the p.d.f. of afterpulses resembles a power-law [13]. In. [14] it is found that the avalanche should be quenched as soon as possible to minimize the avalanche current to minimize filling of the traps, and thus the afterpulsing. In our circuit, prompt quenching is accomplished using a fast comparator, while the series resistor $R_{\mathrm{S}}$ limits the avalanche current. We find that the afterpulsing probability drops from 5.5% for negligible series resistance ($R_{\mathrm{S}} \leq 800\ \Omega$), for which the current is limited by the SPAD itself, to 3.2% for $R_{\mathrm{S}} = 3.3\ \mathrm{k}\Omega$. We find that the afterpulsing probability cannot be lowered below 2.7% even for a very large value of $R_{\mathrm{S}}$ because the junction and parasitic capacitances of the SAP500-T8 already contain enough charge to fill the traps to that level. On the other hand, enlarging $R_{\mathrm{S}}$ beyond 3.3 k$\Omega$ affects the slew rate and worsens the jitter performance, so we take this value as an optimum for our circuit.

## 5. Application to QKD

Our main motivation for an improved single-photon detector is its use in our fast quantum key distribution (QKD) system [15] in which two parties, Alice and Bob, share pairs of hyper-entangled photons [16] generated by spontaneous parametric downconversion (SPDC) in a pair of nonlinear optical BiBO crystals. Photons in a pair are simultaneously entangled in polarization, spatial mode, and time-bin degrees of freedom (DOF). The experimental setup is illustrated in Fig. 12 for one spatial DOF.

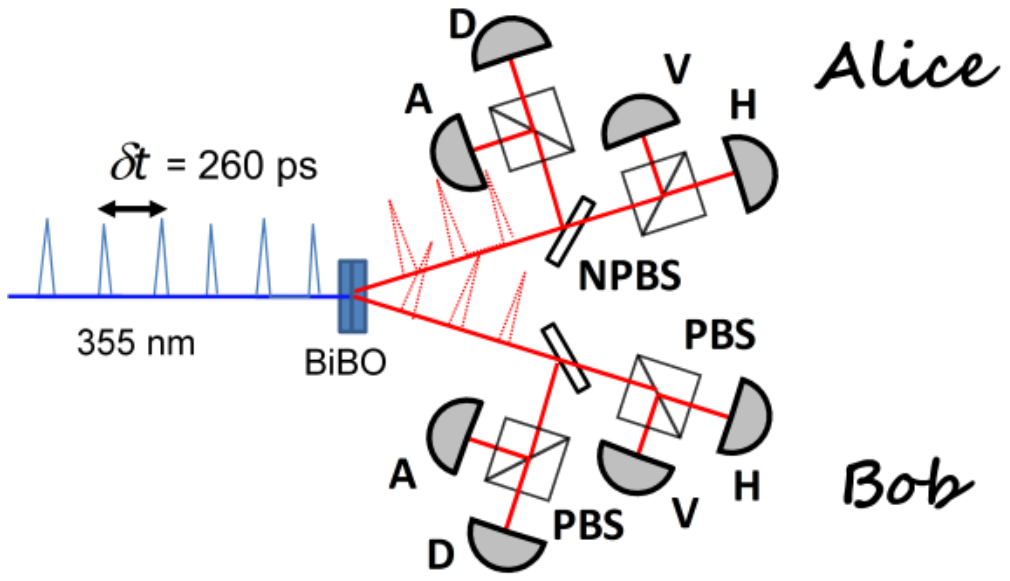

**Fig. 12.** Experimental setup for our QKD system for one spatial mode. The non-polarizing beam splitter (NPBS) in Alice and Bob's setup randomly direct the photonic states to either the Horizontal (H)/Vertical (V) basis or the Diagonal (D)/Anti-Diagonal (A) polarization basis, where single photons are detected and their arrival times recorded.

Each DOF plays a different role in the overall QKD protocol: most of the raw key bits are encoded in the photon timing ("pulse position modulation" allows for up to $\log_2 N$ bits per photon, where $N$ is the number of time bins that constitute a temporal "frame"), polarization entanglement is used to check for eavesdropping, and the spatial modes realize independent quantum communication channels whose purpose is to enhance the secret key rate. The downconversion crystals which generate a maximally entangled polarization state, are pumped by a 355-nm wavelength mode-locked diagonally-polarized (D) pump laser with a 5-ps long pulse duration, and a repetition rate of 120 MHz (8.33-ns pulse period). A repetition-rate multiplication scheme incorporating delay interferometers is used to increase the laser pulse rate by various factors, up to a maximum of a factor of 32 (3.84 GHz rate, 260-ps pulse period). The intensity of the pump laser is adjusted so that Alice and Bob each receive on average 1 photon in a time frame consisting of 1024 successive time bins each 260 ps wide. The polarization portion of the protocol is secured with two mutually-unbiased bases; the details of the security of the time-bin DOF are omitted for brevity and will be discussed in a later paper.

There are several detector-sensitive aspects of this protocol. First, photons are arriving in narrow time slots (bins) that are only 260 ps wide. To capitalize on the high entanglement purity in the time-bin DOF, each detector must have a jitter and detection delay shift (when combined) that is substantially smaller than the bin width. Second, because the time bins are much shorter than the twilight zone of a detector, some of the photons will necessarily arrive in the twilight zone. Such photons, if detected, will suffer a shift of at least a few nanoseconds, which means that the corresponding detection event will appear in a wrong time bin, directly contributing to the bit error rate (BER). Because the detection rate is high (~1 Mcps per detector), a substantial (few percent) portion of the photons will hit the detector during its dead time and immediately after it. Time shifts associated with twilighting and post-dead-time recovery will thus play an important role in the BER and in limiting the maximum achievable secret key rate. Interestingly, due to high rate of entangled photon pair detections by Alice and Bob, dark counts are much suppressed as they contribute minimally to the signal-to-noise ratio (SNR) of the system and do not pose a problem, in contrast to the previous example of the beam splitter RNG. The overall secret key rate for the system shown in Fig. 12 is given by

$$R = M\eta^2 \left(\frac{\langle n \rangle \xi}{\delta t}\right), \tag{4}$$

where $M$ is the number of independent communication channels ($M = 1$ in Fig. 12), $\eta$ is the total efficiency of the channel (assumed the same for Alice and Bob) and includes the spatial collection efficiency of the optics, spectral efficiency of the filters, dead time loss, other losses, and the quantum efficiency of the detectors, $\langle n \rangle$ is the mean generated photon number per time bin, and $\xi$ is the photon efficiency in bits per generated coincidence, which includes the mutual information in the photon arrival time and the polarization. Here, $\xi$ includes the efficiency of the sifting on polarization bases, the error correction efficiency for both the timing and polarization mutual information, and privacy amplification due to leakage of information to an eavesdropper. While $\eta$ is directly proportional to the dead time loss, $\xi$ depends in a complicated way on other imperfections including twilighting, distinguishability and heralding efficiency.

We estimate a raw key rate of around 2.6 Mbit/s when using the SPCM-AQRH detector modules for the case when the pump-laser pulse-repetition-rate is increased by a multiplication factor of 16 (1.92-GHz rate, 521-ps period). At higher repetition rates, the raw key rate decreases because the detector jitter is greater than the bin width. However, with our improved detectors, we can still distinguish pulses at a rate of 3.84 GHz, allowing the system to achieve a raw key rate of 3.6 Mbit/s. The difference between the two detectors in the autocorrelation function $g^2(0)$ at 1.92 GHz is illustrated in Fig. 13. These measurements were made by time-to-digital converter (Keysight Acqiris model U1051A) featuring 50-ps timing resolution. In greater detail, we determine the autocorrelation function for the probability of photon detection. For the pulsed source used here, this correlation should show peaks at the pulse period with a modulation of 100%. From Fig. 13(a), it is seen that the correlation function is nearly featureless for the SPCM-AQRH detector, whereas our custom-built detector shows peaks at the expected location with reasonable modulation depth, as discussed below.

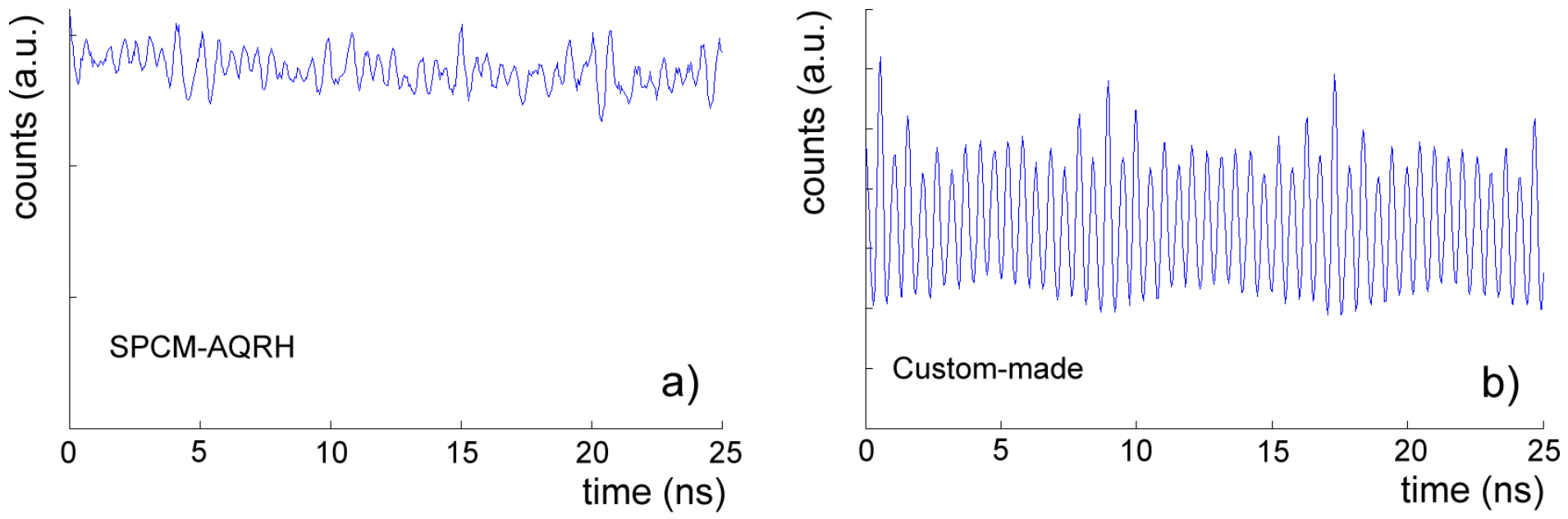

**Fig. 13.** Autocorrelation plots for the a) SPCM-AQRH detector and b) for the custom-made detector. These plots are made using a pump repetition rate of 1.92 GHz, where we can still distinguish the pulses with both detectors. However, the poorer jitter of the SPCM clearly broadens the autocorrelation function.

We further characterize the QKD system and the impact of the detector performance by measuring the distinguishability, heralding efficiency, and cross-correlation of the signals between Alice and Bob. Here, the coincidence window is equal to the bin width, that is $1/f_r$ where $f_r$ is the laser pulse repetition rate.

The distinguishability $\epsilon_H$, defined as the visibility of the autocorrelation plot given in Fig. 13 (having a theoretical maximum of 1), as a function of the laser pulse repetition rate, is shown in Fig. 14(a). We see a fast drop in $\epsilon_H$ as $f_r$ increases for the SPCM-AQRH detector module for bin widths shorter than about 500 ps (*i.e.*, shorter than its effective time resolution), while the custom-made detector exhibits a substantially higher $\epsilon_H$ and no dramatic drop over the measurement range.

The heralding efficiency $\epsilon_H$, defined as the ratio of coincidences and total singles rate (having a theoretical maximum of 1/2), as a function of the detector count rate for $f_r = 1.92$ GHz is shown in Fig. 14(b). In this case, even though the commercial detector SPD-050 has better time resolution than the custom-made detector, due to its longer dead time and much greater twilight zone, displayed in Fig. 5(a), our custom-made detector outperforms it.

The comparison of the custom-made and SPCM-AQRH detectors with respect to losses in the communication channel between Alice and Bob is illustrated by the cross-correlation plots shown in Fig. 14(c). We see a strong twilighting peak after the dead time and larger dead time loss in the SPCM-AQRH when compared to the custom-made detector. The excess number of events appearing in the twilight peak is comparable with the number of events expected to happen in the twilight zone given its width and detection efficiency profile. Both effects cause the observed lower pulse distinguishability and lower secret key rate obtained with this detector.

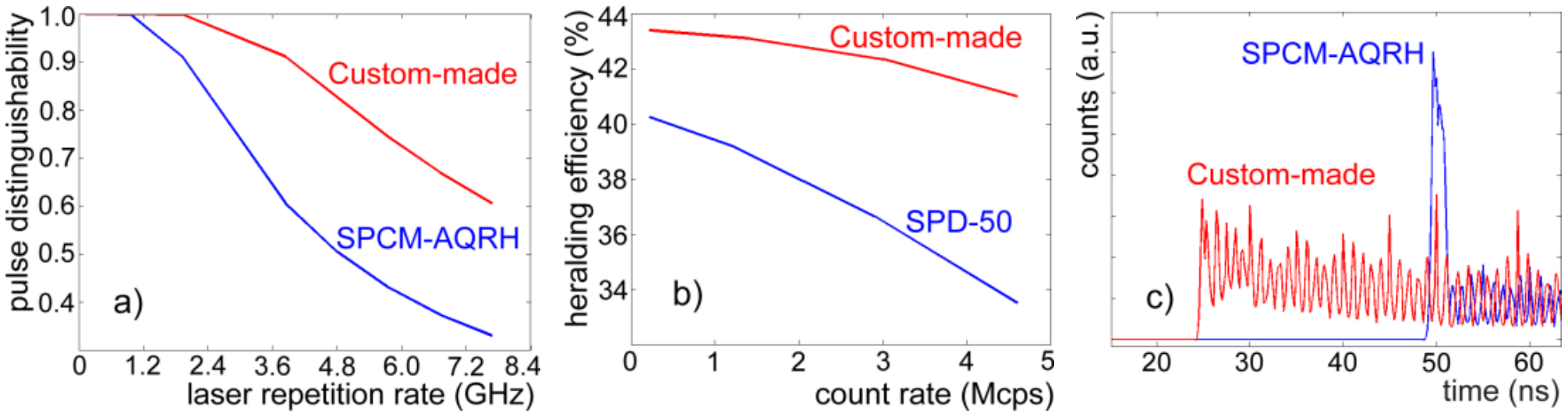

**Fig. 14.** Comparison of various performance metrics of the custom-made and commercial detectors. a) pulse distinguishability as a function of laser pulse repetition rate, where the detection rate is fixed to 4 Mcps by adjusting the detector-beam coupling, b) heralding efficiency, as a function of the detector count rate, for a fixed laser pulse rate of 1.92 GHz, and c) cross-corelation plot for a fixed laser pulse rate of 0.96 GHz.

Our analysis clearly indicates improved performance of our high-dimensional time-bin-based QKD protocol. It is likely that other QKD protocols will similarly benefit from the improved detector performance, especially the stable jitter and reduced deadtime [17, 18]. Also, systems such as optical quantum random number generation based on photon counting [19, 20] will have improved randomness using detectors with lower after pulsing and deadtime [21].

## 6. Conclusions

We report a detailed analysis of imperfections in actively quenched, single-photon avalanche photodiodes operating in Geiger mode. We are particularly interested in imperfections that degrade the performance of systems used for quantum key distribution (QKD). We identify a set of non-standard imperfections, namely, rate-dependence of the dead time, rate-dependence of the jitter, rate-dependence of the detection delay, dead time proximity detection delay shift, twilighting, and jitter degradation. We show that these non-standard imperfections may have a profound effect on the information leakage to an eavesdropper and secure key rate in our and other QKD systems. To minimize these and other imperfections, we designed and constructed a custom-built detector module based on a novel active quenching circuit. Our measurements indicate a great improvement in our QKD system when compared to measurements performed with commercial detectors.

## Funding

We gratefully acknowledge the financial support of the DARPA Defense Sciences Office InPho program, the Office of Naval Research (N00014-13-1-0627) under the Fundamental Research on Frequency-Agile, High-Rate Quantum Key Distribution in a Marine Environment MURI program, and the Ministry of Science Education and Sports of Republic of Croatia (533-19-14-0002).